\documentclass[a4paper,12pt]{article}
	\usepackage[english]{babel}
	\usepackage{blindtext}
	\usepackage{latexsym}
	\usepackage{amscd}
	\usepackage{amssymb}
	\usepackage{amstext}
	\usepackage{amsmath}
	\usepackage{amsxtra}
	\usepackage{fancyhdr}
	\usepackage{graphicx}
	\usepackage{multirow}
\usepackage{bm,amsmath,comment}
	\usepackage{multicol}
	\usepackage{color}
	\usepackage{array}

\begin{document}
	\fontsize{12pt}{22pt}\selectfont
\begin{center}	
\fontsize{16pt}{16pt} \textbf{Highlights of ``A simulation-based study of Zero-inflated Bernoulli model with various models for the susceptible probability"
}

\vskip 3mm
\noindent
\fontsize{14pt}{14pt}\textbf{Essoham Ali}\\
Univ Bretagne Sud, CNRS UMR 6205, Laboratoire de Mathématiques de Bretagne Atlantique, France\\
\vskip 3mm
\noindent
\fontsize{14pt}{14pt}\textbf{Kim-Hung Pho}\\
Fractional Calculus, Optimization and Algebra Research Group, Faculty of Mathematics and Statistics, Ton Duc Thang University, Ho Chi Minh City, Vietnam\\
\vspace{.2in}
Corresponding author: \textcolor{blue}{\textbf{\underline{phokimhung@tdtu.edu.vn}}}
\end{center}	
\vspace{.2in}

\begin{enumerate}
	\item 
This study proposes some new models for family Zero-inflated Bernoulli models.
	\item 
This is the first work to be carried out for this model.
	\item 
Maximum likelihood estimator is used to check its performance of models under consideration.
	\item 
Many simulations and a real data set are performed in this research.
\end{enumerate}

\clearpage	
\begin{center}	
\fontsize{16pt}{16pt} \textbf{A simulation-based study of Zero-inflated Bernoulli model with various models for the susceptible probability
}
\end{center}	
\vspace{.2in}

\begin{center}	
\fontsize{14pt}{16pt}
\textbf{Abstract}
\end{center}		
In this work, we are interested in the stability and robustness of the parameter estimation in the Zero - Inflated Bernoulli (ZIBer) model, when the susceptible probability (SP) model is modeled by numerous different binary models: logit, probit, cloglog and generalized extreme value (GEV). To address this problem, we propose the maximum likelihood estimation (MLE) method to check its performance when different SP models are considered. Based on numerical evidences through simulation studies and the analysis of a real data set, it can be seen that the MLE approach has provided accurate and reliable inferences. In addition, it can also be seen that for the empirical analysis, the probit-ZIBer model is probably more suitable for the fishing data set than the other models considered in this study. Besides, the results obtained in the experimental analysis are also very consistent, compatible and very meaningful in practice. It will help us to understand the importance of increasing production while fishing.

\vspace{.2in}

\vspace{.2in}
\noindent

\clearpage
\section{Introduction}
Statistical science is the science of collecting, analyzing, interpreting and presenting data in order to find out the nature of economic, natural and social phenomena. Statistics allow the summary and easy presentation of numerical information, testing a hypothesis or prediction about the likelihood of an event occurring, and so on. All of these roles are encapsulated in the regression problem. The main aim of this problem is to establish a relationship between an object of interest (response variable) and related objects (predictor variables) to make conclusions with statistical significance.

A linear model is one of the most momentous and ubiquitous statistical models to elucidate relationships between a continuous outcome variable and covariates that could be continuous or discrete. Notwithstanding, in some situations, the outcome variable is not a continuous variable, but a binary variable including yes/no, violated/non-violated, dead/alive, and so on. Thus this model is no longer pertinent for researching the binary data set. To address this issue, Cox (1958) pioneered a logistic model which is now predominantly employed in literature. Up to the present moment, a logistic model has come to be a dominant utensil to scrutinize relationships between the binary dependent variable and covariates.

In practice, when studying the binary data set, it is very common to observe that the phenomenon where the binary reponse (outcome) data have numerous more zeros than expected under a traditional logistic model (LM). Notwithstanding the use of LM in regression analysis, the estimation results for the parameters of interest can be good, their standard errors and interested values may be underestimated. Hence a ZIBer model is a substitute approach to the LM to ameliorate this issue. As we know, this model was first launched by Diop, Diop, and Dupuy (2011) and it is extensively executed and applied in literature.

Some outstanding studies about this topic can be found in 
Diop, Diop, and Dupuy (2011) constructed the asymptotic results of the MLE method for this model.  
Diop, Diop, and Dupuy (2016) performed simulations for the ZIBer model to deduce the important properties for this model. Recently, Li and Lu (2021) have introduced semi-parametric estimators and applications of the ZIBer model. Lee, Pho and Li (2021) proposed the semi-parametric validation likelihood approach to evaluate parameters of this model where the data set contains incomplete values, and goodness of fit test for the ZIBer model is provided by Pho (2022).

Besides, we should note that the ZIBer model is a peculiar case of the zero-inflated (ZI) Binomial (ZIB) model. To the best of our knowledge, the ZIB model is one of the most widespread models which is used to address the excess zeros. The theory and property of ZIB model is contained in Hall (2000). For literature on this, see, e.g., 
He et al. (2015), Rakitzis, Maravelakis, and Castagliola (2016), Diallo, Diop, and Dupuy (2017),  Bodromurti, Notodiputro, and Kurnia (2018), Diallo, Diop, and Dupuy (2019), and Wanitjirattikal and Shi (2020), and so on. 

Analogous to Ali (2022), the author has considered the susceptible probability (SP) model is modeled with several different link models for the ZI Poisson model. This model is also considered the most ubiquitous and crucial regression model in the family of ZI count models (see Truong et al., 2021). Meanwhile, in the family of ZI binary models, ZIBer model is considered as the most popular and important regression model in this family, and hence we aim to research more perspectives about this ZIBer model. In this study, we are interested in a novel issue for the ZIBer model, specifically, this work studies via simulation studies the properties of MLE in the ZIBer models when the SP model is modeled by many different binary models. Our results are also illustrated through an actual data set.

The progress of this work is arranged as follows.
Section 2 shortly describes the sufficient recipe and the related functions of the ZIBer model. The estimating equation of this model is introduced in this section. A variety of the SP models for the ZIBer model is offered in Section 3. Section 4 provides numerous simulation studies. 
Section 5 applies the fishing data set to the ZIBer model. 
Some of crucial discussions are illustrated in the next section. The final section is destined for the conclusion of the paper.

\section{Model and Estimating Equations}
\subsection{Zero-Inflated Bernoulli (ZIBer) model}
It is assumed that $Y_i=1$ if the $i$-th individual has an event of interest and $Y_i=0$ otherwise.
$X_i$ and $Z_i$ be covariates. Let $W_i$ be an implicit binary variable showing whether or not an individual is susceptible to the event of concern, e.g., $W_i=1$ if susceptible; $W_i=0$ otherwise. The value of $W_i$ is unknown in the situation $Y_i=0$. We consider the following model for the SP of the subject (model A):
\begin{align*}
\log \left(\frac{P(W_i= 1|Z_i)}{1 - P(W_i=1|Z_i)} \right) = \gamma_0 + \gamma_1^\top Z_i=\bm{\gamma}^\top\mathcal Z_i,
\end{align*}
where  $\mathcal Z_i=(1, Z_i^\top)^\top$,
and the probability of occurrence of the event for the susceptible individual (model B):
\begin{align*}
\left\{
\small
\begin{array}{ll}
\log \left(\dfrac{P(Y_i=1|X_i,Z_i, W_i=1)}{1 - P(Y_i=1|X_i,Z_i, W_i=1)} \right) = \eta_0 +\eta_1^\top{X_i} + \eta_2^\top{Z_i}=\bm\eta^\top\mathcal X_i & 
\\
P(Y_i=1|X_i,Z_i, W_i=0)=0, & 
\end{array}
\right.
\end{align*}
where $\mathcal {X}_i = (1, X_i^\top, Z_i^\top)^\top$. The ZIBer model is given below (see Lee et al., 2021):
\begin{align}
\label{original.function}
P(Y_i=1|X_i,Z_i)=
\frac{e^{\bm\gamma^\top\mathcal Z_i}}{1+e^{\bm\gamma^\top\mathcal Z_i}}\frac{e^{\bm\eta^\top\mathcal X_i}}
{1+e^{\bm\eta^\top \mathcal {X}_i}}=H(\bm\gamma^\top\mathcal Z_i)H(\bm\eta^\top\mathcal {X}_i) ,
\end{align}
where $H(t)=[1+\exp(-t)]^{-1}$, $\bm\beta=(\bm\gamma^\top,\bm\eta^\top)^\top$ with $\bm\gamma=(\gamma_0,\gamma_1^\top)^\top$ and  $\bm\eta=(\eta_0,\eta_1^\top,\eta_2^\top)^\top$
to be evaluated.

The likelihood function of the sample $\{(Y_i,\mathcal{X}_i):i=1,\dots,n\}$ is described as:
\begin{align*} 
L(\bm\beta)= \prod\limits_{i = 1}^n \left\{ \left[ \frac{e^{\bm\gamma^\top\mathcal Z_i+\bm\eta^\top \mathcal {X}_i}}{(1 + e^{\bm\gamma^\top\mathcal Z_i})(1 + e^{\bm\eta^\top\mathcal {X}_i})} \right]^{Y_i} \left[ 1 - \frac{e^{\bm\gamma^\top\mathcal Z_i+\bm\eta^\top \mathcal X_i}}{(1 + e^{\bm\gamma^\top\mathcal Z_i})(1 + e^{\bm\eta^\top\mathcal X_i})} \right]^{1 - Y_i} \right\}.
\end{align*}
The log-likelihood function
$\ell(\bm\beta)=\log [L(\bm\beta)]$ is:
\begin{align} 
\label{log.likelihood}
\ell(\bm\beta)=\log [L(\bm\beta)]
&=\sum\limits_{i = 1}^n Y_i ( \bm\gamma^\top\mathcal Z_i + \bm\eta^\top\mathcal X_i) 
+ \sum\limits_{i = 1}^n (1-Y_i)\left[\log (1 + e^{\bm\gamma^\top\mathcal Z_i}+e^{\bm\eta^\top\mathcal X_i}) \right] \\ \nonumber
& \hskip 4mm
-\sum\limits_{i = 1}^n \left[\log ( 1 + e^{\bm\gamma^\top\mathcal Z_i})\right]- \sum\limits_{i = 1}^n \left[\log(1 + e^{\bm\eta^\top\mathcal X_i}) \right] .
\end{align}

\subsection{Estimating Equations for Complete Data}
According to $\ell(\bm{\beta})$, where $\ell(\bm\beta)$ is illustrated in (\ref{log.likelihood}),  the score
function is depicted below:
\begin{align}
\label{MLE.function}
U_{F,n}(\bm\beta)
=\dfrac{1}{\sqrt{n}}\dfrac{\partial\ell(\bm\beta)}{\partial\bm\beta}
=\dfrac{1}{\sqrt{n}}
\begin{pmatrix}
\dfrac{\partial\ell(\bm\beta)}{\partial\bm\gamma} \\
\vspace*{-0.4cm}\\
\dfrac{\partial\ell(\bm\beta)}{\partial\bm\eta}
\end{pmatrix}
=\dfrac{1}{\sqrt{n}}\sum_{i=1}^nD_i(\bm\beta).
\end{align}
Here $D_i(\bm\beta)=\partial\ell_i(\bm\beta)/\partial\bm\beta
=\left(D_{i1}^\top\left(\bm\beta\right),D_{i2}^\top\left(\bm\beta\right)\right)^\top$,
$i=1,\dots,n$,
\begin{align}
D_{i1}(\bm\beta)
&=\dfrac{\partial\ell_i(\bm\beta)}{\partial\bm\gamma}
=\mathcal{Z}_iB_{i1}(\bm\beta)[Y_i-H(\bm\gamma^\top\mathcal{Z}_i)H(\bm\eta^\top\mathcal{X}_i)],\label{score1} \\
D_{i2}(\bm\beta)
&=\dfrac{\partial\ell_i(\bm\beta)}{\partial\bm\eta}
=\mathcal{X}_iB_{i2}(\bm\beta)[Y_i-H(\bm\gamma^\top\mathcal{Z}_i)H(\bm\eta^\top
\mathcal{X}_i)],\label{score2}
\end{align}
for
\begin{align*}
B_{i1}(\bm\beta)=\dfrac{1+e^{\bm\eta^\top\mathcal{X}_i}}{1+e^{\bm\gamma^\top\mathcal{Z}_i}+e^{\bm\eta^\top\mathcal{X}_i}}
\quad\textrm{and}\quad
B_{i2}(\bm\beta)=\dfrac{1+e^{\bm\gamma^\top\mathcal{Z}_i}}{1+e^{\bm\gamma^\top\mathcal{Z}_i}+e^{\bm\eta^\top\mathcal{X}_i}}.
\end{align*}
Let $\mathcal{T}_i=(\mathcal{Z}_i^\top,\mathcal{X}_i^\top)^\top$ and
\begin{align}
\label{Ai}
B_i(\bm\beta)
=\begin{pmatrix}
B_{i1}(\bm\beta)\bm{I}_{(b+1)} &\bm{0}_{(b+1)\times(a+b+1)} \\
\bm{0}_{(a+b+1)\times(b+1)}& B_{i2}(\bm\beta)\bm{I}_{(a+b+1)}
\end{pmatrix},
\end{align}
where $\bm{I}_{(b+1)}$ and $\bm{I}_{(a+b+1)}$ are
$(b+1)\times(b+1)$ and $(a+b+1)\times(a+b+1)$ identity matrices,
respectively. $D_i(\bm\beta)$ can then be re-written as
\begin{align*}
D_i(\bm\beta)
=B_i(\bm\beta)\mathcal{T}_i[Y_i-H(\bm\gamma^\top\mathcal{Z}_i)H(\bm\eta^\top\mathcal{X}_i)].
\end{align*}
Note that
\begin{align*}
E[D_i(\bm\beta)]
&=E[E(D_i(\bm\beta)|X_i,Z_i)]\\
&=E\{E[B_i(\bm\beta)\mathcal{T}_i(Y_i-H(\bm\gamma^\top\mathcal{Z}_i)H(\bm\eta^\top\mathcal{X}_i))|X_i,Z_i] \}\\
&= E\{B_i(\bm\beta)\mathcal{T}_iE[(Y_i -
H(\bm\gamma^\top\mathcal{Z}_i)H(\bm\eta^\top\mathcal{X}_i))|X_i,Z_i]\}=0
\end{align*}
Because $E[U_{F,n}(\bm\beta)]=0$,
$U_{F,n}(\bm\beta)$ is an unbiased score function. The MLE $\widehat{\bm\beta}_F$ of $\bm\beta $
can be obtained by addressing $U_{F,n}(\bm\beta)=0$. It should also be said that the asymptotic properties of the estimator of $\bm\beta$ has been brilliantly demonstrated in the studies of Lee, Pho and Li (2021) and Pho (2022). To save space, the issues are not discussed further here.
\section{A variety of the SP models for the ZIBer model}
In this work, we aim to describe the general formula for the ZIBer model when the SP model is modeled by numerous different binary models.
\subsection{The logit-ZIBer model}
We should also note that in Section 2.1, model A is simulated by a logit model, and the formula (\ref{original.function}) is also the general formula of the ZIBer model in this situation. If we would like to call it the most complete name, then it can be called the logit-ZIBer model. Besides, we named it so for the purpose of making it easy to distinguish from other ZIBer model when the SP model is modeled by numerous different binary models.
\subsection{The probit-ZIBer model}
In this situation, the formula of model A will change, and the formula of model B will not change. Specifically, the formula of model A is built as follows:
\begin{align*}
P(W_i= 1|Z_i)= \Phi \left(\gamma+\gamma_1 Z_{i1}+\cdots+\gamma_p Z_{ip}\right) := \Phi \left(\bm{\gamma}^\top\mathcal Z_i\right),
\end{align*}
where $\Phi$ is the cumulative distribution function (CDF) of the $N(0,1)$, and then the general formula of the probit-ZIBer model will become:
\begin{align}
\label{original.function.probit}
P(Y_i=1|X_i,Z_i)=
\Phi(\bm\gamma^\top\mathcal Z_i)H(\bm\eta^\top\mathcal {X}_i) ,
\end{align}

The log-likelihood function of Probit-ZIBer model is:
\begin{eqnarray*}\label{loglikelihood.probit}
\ell(\bm\beta) &=& \sum_{i=1}^n \left\{Y_i\left[\bm\eta^\top\mathcal {X}_i+\log\Phi \left(\bm{\gamma}^\top\mathcal Z_i\right)\right] +(1-Y_i)\log\left[1+\left(1-\Phi (\bm{\gamma}^\top\mathcal Z_i)\right)e^{\bm\eta^\top\mathcal {X}_i}\right] \right. \nonumber \\
&&\hspace{2cm} \left.-\log(1+e^{\eta^\top\mathcal X_i})
\right\}.
\end{eqnarray*}

\subsection{The cloglog-ZIBer model}
Similar to above, in this scenario, the formula of model A will also change, while the formula of model B will not change. Explicitly, the formula for model A is set up as follows:
\begin{align*}
-\log\left(-\log\left(P(W_i= 1|Z_i)\right)\right) 	 =\bm{\gamma}^\top\mathcal Z_i,
\end{align*}
This would be equivalent to
\begin{align*}
P(W_i= 1|Z_i) = e^{-\exp\left(-\bm{\gamma}^\top\mathcal Z_i\right)},
\end{align*}
and then the general formula of the cloglog-ZIBer model will have the form:
\begin{align}\label{original.function.cloglog}
P(Y_i=1|X_i,Z_i)=e^{-\exp\left(-\bm{\gamma}^\top\mathcal Z_i\right)}
H(\bm\eta^\top\mathcal {X}_i).
\end{align}
The likelihood function of $\bm\beta$ is
calculated as
$\ell(\bm\beta)=\log [L(\bm\beta)]$ where:
\begin{eqnarray*}
\label{likelihood.cloglog}
L(\bm\beta)= \prod\limits_{i = 1}^n \left\{ \left[ e^{-\exp\left(-\bm{\gamma}^\top\mathcal Z_i\right)}H(\bm\eta^\top\mathcal {X}_i) \right]^{Y_i} \left[ 1 - e^{-\exp\left(-\bm{\gamma}^\top\mathcal Z_i\right)}H(\bm\eta^\top\mathcal {X}_i) \right]^{1 - Y_i} \right\},
\end{eqnarray*}
from which we easily deduce the loglikelihood
\begin{eqnarray*}\label{loglikelihood.cloglog}
\ell(\bm\beta) &=& \sum_{i=1}^n \left\{Y_i\left[\bm\eta^\top\mathcal {X}_i-
 e^{-\exp\left(-\bm{\gamma}^\top\mathcal Z_i\right)}\right] +(1-Y_i)\log\left[1+\left(1-e^{-\exp\left(-\bm{\gamma}^\top\mathcal Z_i\right)}\right)e^{\bm\eta^\top\mathcal {X}_i}\right] \right. \nonumber \\
&&\hspace{2cm} \left.-\log(1+e^{\eta^\top\mathcal X_i})
\right\}.
\end{eqnarray*}

\subsection{The GEV-ZIBer model}
A key component of the count data is the specification of the link function.  The usual binary regression model may not be appropriate because of the potential non-linearity in the latent regression function and the assumed skewness in a given link function, such as the symmetric probit and the asymmetric cloglog links.
Limiting the latent regression function to a simple linear or parametric form is clearly restrictive in modeling the binary data. Moreover, mis-specification in the link function leads to an increase in mean square error of the estimated probability and substantial bias in estimating the regression parameters.
Inspired by the work of Wang and Dey (2010), in this section we propose to use a link function based on the GEV distribution. The GEV distribution function $\text{GEV}(\mu, \sigma, \epsilon)$ is given by:
\begin{eqnarray*}\label{GEV-ZIBer}
F(x|\mu,\sigma,\epsilon)=  \left\{
\begin{array}{ll}
 \exp\left[ -\left\lbrace 1+\epsilon\frac{(x-\mu)}{\sigma}\right\rbrace_+ ^{-\frac{1}{\epsilon}}\right], &  \epsilon\neq 0, \\
 
\hspace{1cm}\exp\left\lbrace -\exp(\frac{(x-\mu)}{\sigma})\right\rbrace, &  \epsilon= 0,
\end{array}
\right.
\end{eqnarray*}
where $\mu\in \mathbb{R} $ is the location parameter, $\sigma \in\mathbb{R}_+$ is the scale parameter and $\epsilon\in\mathbb{R}$ is the
shape parameter and $x_+ = \text{max}(0, x)$.  The importance of using the GEV as a link function arises from the fact that the shape parameter $\epsilon$ purely controls the behavior of the tail of the distribution.

For a binary response variable $W_i$ and the vector of explanatory variables $\mathcal Z_i$, let $\omega(z_i)=P(W_i= 1|Z_i=z_i))$.  The link function of the GEV model is given by
\begin{eqnarray}
\frac{1-\left[ \mbox{log}(1-\omega (z_i))\right]^{-\epsilon}}{\epsilon}=\gamma^\top\mathcal Z_i.
\end{eqnarray}
This will lead to 
\begin{eqnarray}
\omega(z_i)=P(W_i= 1|Z_i=z_i)=1-\exp\left\lbrace \left[(1-\epsilon\gamma^\top\mathcal Z_i)_+\right]^{-\frac{1}{\epsilon}}\right\rbrace= 1-\mbox{GEV}(-\gamma^\top\mathcal Z_i,\epsilon)
\end{eqnarray}
where $\mbox{GEV}( z; \epsilon)$ represents the cumulative probability at $z$ for the generalized extreme
value distribution with $\mu = 0, \; \sigma = 1$, and an unknown shape parameter $\epsilon$.
A simple calculation yields the following general formula for the GEV-ZIBer model:
\begin{eqnarray}
P(Y_i=1|X_i,Z_i) & = & \omega(z_i)H(\bm\eta^\top\mathcal {X}_i) \\ \nonumber
& = & (1-\mbox{GEV}(-\gamma^\top z_i,\epsilon))\frac{e^{\bm\eta^\top\mathcal X_i}}{1+e^{\bm\eta^\top \mathcal {X}_i}}.
\end{eqnarray}

Let $\bm\beta=(\bm\gamma^\top,\bm\eta^\top,\bm\epsilon)^\top$ denote the unknown $k$-dimensional ($k = p + q + 1$) parameter in
the conditional distribution of $Y_i$ given $\mathcal X_i$ and $\mathcal Z_i$. The the log-likelihood function is as follows
\begin{eqnarray*}\label{likelihood.GEV}
\ell(\bm\beta) &=& \sum_{i=1}^n \left\{Y_i\left[\bm\eta^\top\mathcal {X}_i+\log\omega(\mathcal Z_i)\right] +(1-Y_i)\log\left[1+\left(1-\omega(\mathcal Z_i\right))e^{\bm\eta^\top\mathcal {X}_i}\right] -\log(1+e^{\eta^\top\mathcal X_i})
\right\}.
\end{eqnarray*}
\section{Simulation studies}
We aim to test two cases in this section, the first being the simplest case that is both $X$ and $Z$ a univariate. Whereas the other situation we consider has many variables in covariates. In the current simulation, the sample sizes were n = 200; 500; 1,000 and 2,000. Comparisons based on bias, asymptotic standard error (ASE), standard deviation (SD) and coverage probability (CP). In order to make it easier for readers to see the difference of the general formulas of the ZIBer model, we all repeat them carefully before conducting the simulation.

\noindent\textit{Case 1: Both $\bm{X}$ and $\bm{Z}$ a univariate.}

\noindent\textbf{Scenario (i)}: We simulate data from a  logit-ZIBer model (model A)  defined by : 
\begin{align*}
P(Y_i=1|X_i,Z_i)=H\left(\gamma_0 + \gamma_1^\top Z_i\right).H\left(\eta_0 +\eta_1^\top{X_i} + \eta_2^\top{Z_i}\right)=H\left(\bm{\gamma}^\top\mathcal Z_i\right).H(\bm\eta^\top\mathcal {X}_i).
\end{align*}
Parameters $\gamma$ and $\eta$ are chosen as $\gamma=(-0.8, 0.9)^\top$ and $\eta=(0.7, -1.7, 0.5)^\top.$

\noindent\textbf{Scenario (ii)}: The general formula of the probit-ZIBer model (model B) is as follows:
\begin{align*}
P(Y_i=1|X_i,Z_i)=\Phi\left(\gamma_0 + \gamma_1^\top Z_i\right).H\left(\eta_0 +\eta_1^\top{X_i} + \eta_2^\top{Z_i}\right)=\Phi\left(\bm{\gamma}^\top\mathcal Z_i\right).H(\bm\eta^\top\mathcal {X}_i) ,
\end{align*}
The regression parameters $\gamma$ and $\eta$  are chosen as follows: $\gamma=(-0.8, 0.9)^\top$ and $\eta=(0.7, -1.7, 0.5)^\top.$

\noindent\textbf{Scenario (iii)}: The simulation design is as follows.
For each of n individuals, the count response $Y$ is simulated from a  clogclog-ZIBer model (model C)  follows
\begin{align*}
P(Y_i=1|X_i,Z_i)=e^{-\exp\left[-\left(\gamma_0 + \gamma_1^\top Z_i\right)\right]}.H\left(\eta_0 +\eta_1^\top{X_i} + \eta_2^\top{Z_i}\right)
=e^{-\exp\left(-\bm{\gamma}^\top\mathcal Z_i\right)}.H(\bm\phi^\top\mathcal {X}_i) ,
\end{align*}
where $\gamma=(0.5, -0.5)^\top$ and $\eta=(-0.5, -1.2, 0.5)^\top$

\noindent\textbf{Scenario (iv)}: The simulation setting is as follows. We consider  the GEV-ZIBer model (model D)  :
\begin{align*}
P(Y_i=1|X_i,Z_i)
&=\left[1-\mbox{GEV}(-\left(\gamma_0 + \gamma_1^\top Z_i\right),\epsilon)\right].H\left(\eta_0 +\eta_1^\top{X_i} + \eta_2^\top{Z_i}\right)\\
&=\left[1-\mbox{GEV}(-\gamma^\top\mathcal Z_i,\epsilon)\right].H(\bm\phi^\top\mathcal {X}_i) ,
\end{align*}
where $\gamma=(-0.5, 0.5)^\top$ and $\eta=(-0.5, -1.5, 0.5)^\top.$

In all four scenarios above, then $\bm\beta=(\bm\gamma^\top,\bm\eta^\top)^\top$ with $\bm\gamma=(\gamma_0,\gamma_1^\top)^\top$ and  $\bm\eta=(\eta_0,\eta_1^\top,\eta_2^\top)^\top$
to be evaluated. The data of X and Z were generated by $\mathcal{N}(0, 1)$ and $\mathcal{B}(1, 0.5)$, respectively. According to the general form of model A, B, C and D to create the data of Y. It should also be mentioned that the ZI ratios in the A, B, C and D models are different, and they take the values of 58\%, 63\%, 87\% and 87\%, respectively.

\noindent\textit{Case 2: Both $X$ and $Z$ are  vectors of covariates}

\noindent\textbf{Scenario (i)}:The general form of the logit-ZIBer model (model A) is as follows:
\begin{align*}
P(Y_i=1|X_i,Z_i)
&=H\left(\gamma_0 + \gamma_1^\top Z_{1i} + \gamma_2^\top Z_{2i}\right).H\left(\eta_0 +\eta_1^\top X_{1i} +\eta_2^\top X_{2i}+ \eta_3^\top X_{3i} +\eta_4^\top Z_{2i}\right)\\
&=H\left(\bm{\gamma}^\top\mathcal Z_i\right).H(\bm\eta^\top\mathcal {X}_i) ,
\end{align*}
 with $(\gamma_0,\gamma_1,\gamma_2)^\top=(0.5,0.2,-0.6)^\top$ and  $(\eta_0,\eta_1,\eta_2,\eta_3,\eta_4)^\top=(0.7,-1.7,0.5, -1.2, 0.5)^\top$

\noindent\textbf{Scenario (ii)}: The general formula of the probit-ZIBer model (model B) is as follows:
\begin{align*}
P(Y_i=1|X_i,Z_i)
&=\Phi\left(\gamma_0 + \gamma_1^T Z_{1i} + \gamma_2^T Z_{2i}\right).H\left(\eta_0 +\eta_1^TX_{1i} +\eta_2^TX_{2i} +\eta_3^TX_{3i} +\eta_4^TZ_{2i}\right)\\
&=\Phi\left(\bm{\gamma}^\top\mathcal Z_i\right).H(\bm\eta^\top \mathcal {X}_i) ,
\end{align*}
 with $(\gamma_0,\gamma_1,\gamma_2)^\top=(0.5,0.2,-0.6)^\top$ and  $(\eta_0,\eta_1,\eta_2,\eta_3,\eta_4)^\top=(0.7,-1.7,0.5, -1.2, 0.5)^\top$.

\noindent\textbf{Scenario (iii)}: The general form of the clogclog-ZIBer model (model C) is as follows:
\begin{align*}
P(Y_i=1|X_i,Z_i)
&=e^{-\exp\left[-\left(\gamma_0 + \gamma_1^\top Z_{1i} + \gamma_2^\top Z_{2i}\right)\right]}.H\left(\eta_0 +\eta_1^\top X_{1i} +\eta_2^\top X_{2i}+\eta_3^\top X_{3i} +\eta_4^\top Z_{2i}\right)\\
&=e^{-\exp\left(-\bm{\gamma}^\top\mathcal Z_i\right)}.H(\bm\phi^\top \mathcal {X}_i) ,
\end{align*}
with $(\gamma_0,\gamma_1,\gamma_2)^\top=(0.5,-0.2,0.5)^\top$ and  $(\eta_0,\eta_1,\eta_2,\eta_3,\eta_4)^\top=(-0.5,0.5,0.5, -0.5, 0.5)^\top$.

\noindent\textbf{Scenario (iv)}: The general formula of the GEV-ZIBer model (model D) is as follows:
\begin{align*}
P(Y_i=1|X_i,Z_i)
&=\left[1-\mbox{GEV}(-\left(\gamma_0 + \gamma_1^\top Z_{1i} + \gamma_2^\top Z_{2i}\right),\epsilon)\right].H\left(\eta_0 +\eta_1^\top X_{1i} +\eta_2^\top X_{2i} +\eta_3^\top X_{3i} +\eta_4^\top Z_{2i}\right)\\
&=\left[1-\mbox{GEV}(-\gamma^\top\mathcal Z_i,\epsilon)\right].H(\bm\phi^T\mathcal {X}_i) ,
\end{align*}
with $(\gamma_0,\gamma_1,\gamma_2)^\top=(0.5,-0.2,0.5)^\top$ and  $(\eta_0,\eta_1,\eta_2,\eta_3,\eta_4)^\top=(0.7,-0.5,0.5,-0.7,0.5)^\top$.

\noindent In all four scenarios above, then $\bm\beta=(\bm\gamma^\top,\bm\eta^\top)^\top$ with $\bm\gamma=(\gamma_0,\gamma_1^\top,\gamma_2^\top)^\top$ and  $\bm\eta=(\eta_0,\eta_1^\top,\eta_2^\top,\eta_3^\top,\eta_4^\top)^\top$
to be evaluated.

\noindent The data of $X_1$, $X_2$ and $X_3$ were generated by a normal N(0, 1), exponential exp(1) and Bernoulli $\mathcal{B}(1, 0.3)$ distributions, respectively. The data of $Z_1$ and $Z_2$ were generated by a  Bernoulli $\mathcal{B}(1, 0.5)$, normal $\mathcal{N}(-1, 1)$ distributions, respectively. Basing on the general formula of model A, B, C and D to create the data of $Y$. It should also be mentioned that the ZI ratios in the A, B, C and D models are the same, and they get the same value, which is 70\%.

 \textit{Results:} For each combination \texttt{[sample size $\times$ proportion of ZI]} we simulate N = 1,000 samples and for each of them, we calculate the maximum likelihood estimate $\widehat{\bm\beta}$ of $(\bm\gamma,\bm\eta)$ (case 1) and $(\bm\gamma,\bm\eta)$ (case 2)
. Simulations are conducted using \texttt{optim} function the statistical software \texttt R.
 
 Based on the N estimates, we obtain, for each simulation scenario, the : i) empirical
bias  ii) asymptotic ASE and empirical SD of each estimator, iii) empirical CP of 95\%-level confidence interval for each parameter. For case 1, results are given in Tables 1 and 2. For case 2, results
are given in Table 3 and 4.

From these tables, the bias of the MLE approach in evaluating the parameters for all four models at all sample sizes are small. It is also easy to see that in the case of the larger sample size in the investigated simulation, the biases are smaller. The next two values of interest are ASE and SD, which are very close to each other. They have an attractive feature that the ASE and SD parameters will receive values that tend to decrease as the surveyed sample population increases.  Besides, the last parameter of concern in this part is that CP is very close to 0.95.

In general, from these tables we can see that, the bias, ASE, and SD  of all estimators decrease as sample size increases. Moreover, the empirical CPs are close to the nominal confidence level.
In the  scenario (iv) of case 1 , the bias estimates increase substantially, resulting in coverage probabilities decrease around 85\%-90\%, which might indicate that the asymptotic variance is slightly under-estimated. 

On the other hand, the bias of the Model C (see Table 4) stays moderate, and is similar to the first scenario of case 1, which is also expected due to the  robustness of the link function. For fixed $n$, we can observe that  performances of the $\bm\gamma$ remain stable when the proportion of ZI varies from small to moderate values  and deteriorate when ZI achieves higher values, and  performances of the $\bm\eta$ improve and then deteriorate as ZI increases. 

 Hence, it can be said that the MLE method seems to provide an efficient and reliable method for estimating  models, even when the number of parameters is quite small.

\clearpage
\begin{table}[h!]
\caption{Simulation results for case that is both $X$ and $Z$ a univariate. 
}  
\label{tab:case1} \centering \tabcolsep=3.5pt 
\begin{tabular}{cl cccc c cccc }
\hline
& & \multicolumn{4}{c}{$n=200$} 
& &\multicolumn{4}{c}{$n=500$} \\
\cline{3-6} \cline{8-11} 
&&Model A &Model B&Model C&Model D
&&
Model A &Model B&Model C&Model D
\\
\hline
$\gamma_0$& Bias& 
0.4790 &  0.4255&  -0.2548& -0.5457 && 0.1023&   0.2424&  -0.1275&-0.3474\\
&ASE& 2.9321&  3.1266 &  2.1732& 5.9064 &&0.7219 &  1.4342&  1.1162&  2.7924\\
&SD& 1.6654&  1.1084& 1.1448 &  1.3580 && 0.7426&  0.8110&   0.7842&1.0271\\
&CP&  0.9340&  0.9540&  0.9420&   0.8710 && 0.9320&  0.9250& 0.8940& 0.9030\\
$\gamma_1$& Bias& -0.4078&  -0.3964&  0.2336&  0.5509 && -0.0627&  -0.2036& 0.1115& 0.3213\\
&ASE& 3.0275& 3.2937&  2.4056& 5.9643 && 0.7653&  1.5244&  1.3154&2.7927\\
&SD& 1.6098&  1.0838&  1.1087& 1.3213 && 0.7386&  0.7520&   0.7750&0.9966\\
&CP&  0.9740& 0.9820&  0.9960& 0.9450 && 0.9720&  0.9840&  0.9850&  0.9530\\
$\eta_0$& Bias& 0.2475&  0.2867& 0.3681&  0.2125&& 0.1036&  0.1729& 0.3348& 0.0475\\
&ASE& 1.0719& 1.6407&  2.1510&  0.6811 &&  0.6059&   1.0821&  1.4603&0.4014\\
&SD& 1.1045&  1.4354&  1.3427& 0.8154
& & 0.6616 &  1.0741&   1.2185& 0.4270\\
&CP& 0.9270&  0.8840&  0.9620& 0.8980
& & 0.9490& 0.9280&  0.9810&0.8730\\
$\eta_1$& Bias& 
-0.4145&  -0.5176&  -0.6072&-0.4092
& & -0.1548&  -0.2443&  -0.4585& -0.1224\\
&ASE& 0.9080&  0.9454&  1.3068&  0.6709
& & 0.4765&  0.5477& 0.8168&0.3458\\
&SD& 0.9331 &  0.9790& 0.9319 &  0.7216
& &  0.5388& 0.5883&  0.8850&0.3639\\
&CP& 0.9580 & 0.9660& 0.9940 &    0.9880
& & 0.9580&   0.9360&  0.9830& 0.9440\\
$\eta_2$& Bias& 
0.1429&   0.3471 & 0.1037 &  0.1050
& &  0.0302&  0.1659&  0.0799&0.0379\\
&ASE&  1.0187&  1.5302& 2.2604 & 0.9414
& &  0.5714&  0.8823&  1.3962&0.5493\\
&SD& 0.9741&  1.2678& 1.3341 & 0.9343
& & 0.6034&  0.8155&  1.1230& 0.5506\\
&CP& 0.9860&  0.9940& 0.9980 &  0.9710
& & 0.9480& 0.9600&  0.9910&  0.9540\\
\hline
\end{tabular}
\end{table}

\clearpage
\begin{table}[h!]
\caption{Simulation results for case that is both $X$ and $Z$ a univariate 
}  
\label{tab:case2} \centering \tabcolsep=3.5pt 
\begin{tabular}{cl cccc c cccc }
\hline
& & \multicolumn{4}{c}{$n=1000$} 
& &\multicolumn{4}{c}{$n=2000$} \\
\cline{3-6} \cline{8-11} 
&&Model A &Model B&Model C&Model D
&&
Model A &Model B&Model C&Model D
\\
\hline
$\gamma_0$& Bias& 
 0.0337&  0.1002&  -0.1108&   -0.2101
& & 0.0188&  0.0381&    -0.0555& -0.0970\\
&ASE& 0.2802&   0.5145& 0.5946 &  1.2490
& & 0.1869&  0.2159&   0.3523& 0.4590\\
&SD& 0.2903&  0.4697&   0.9070&  0.7146
& & 0.1963&  0.2308&    0.3985& 0.4316\\
&CP& 0.9510& 0.9190& 0.9070& 0.9300
& & 0.9540&   0.9280&  0.9150& 0.9430\\
$\gamma_1$& Bias& 
-0.0039&  -0.0385&  0.0610&  0.1829
& &  0.0002&  0.0049&  0.0224&0.0733\\
&ASE& 0.3098&  0.5739&   0.8662&   1.2326
& & 0.2053&   0.2182&  0.3756&0.4439\\
&SD& 0.3162&  0.4136& 0.5704 & 0.6864
& &  0.2064&  0.2073&   0.3755& 0.3998\\
&CP& 0.9680&  0.9800&  0.9790& 0.9580
& &  0.9580&  0.9780&   0.9800& 0.9580\\
$\eta_0$& Bias& 
 0.0450&   0.0788&    0.0851 & -0.0021
& & 0.0188&  0.0319&   0.0559& -0.0076\\
&ASE& 0.4063& 0.7613&  1.0479&   0.2874
& & 0.2800&   0.5167&  0.6903&  0.2042\\
&SD&  0.4200&  0.7671&  0.9121&  0.2886
& & 0.2917&  0.5325&   0.6896&  0.2183\\
&CP&0.9650&  0.9550&  0.9770&   0.9130
& &  0.9510&   0.9700&   0.9740&  0.9220\\
$\eta_1$& Bias& 
 -0.0503&  -0.1036&  -0.1877&-0.0351
& & -0.0292&  -0.0462& -0.0977& 0.9220\\
&ASE&  0.3105&   0.3524&  0.4781& 0.2334
& & 0.2148&  0.2380&   0.3128&  0.1633\\
&SD& 0.3246&  0.3812& 0.5059&  0.2416
& &  0.2275&  0.2424&  0.3292& 0.1658\\
&CP& 0.9460&  0.9510&  0.9710&  0.9320
& &  0.2275&  0.9530&    0.9610& 0.9440\\
$\eta_2$& Bias& 
-0.0002&  0.0297 &   0.0748&  0.0252
& & 0.0110&  0.0091&  0.0136& 0.0071\\
&ASE& 0.3888&   0.5919&     0.9686& 0.3782
& &  0.2698&   0.4018&  0.6365&0.2629\\
&SD& 0.3959&  0.6046&    0.8728& 0.3955
& & 0.2712&   0.4013&  0.6131&0.2749\\
&CP&  0.9460&  0.9520& 0.9690& 0.9710
& &  0.9530&  0.9560&   0.9550& 0.9450\\
\hline
\end{tabular}
\end{table}

\clearpage
\begin{table}[h!]
\caption{Simulation results for the case is both $X$ and $Z$ are  vectors of covariates}  
\label{tab:case3} \centering \tabcolsep=3.5pt 
\begin{tabular}{cl cccc c cccc }
\hline
& & \multicolumn{4}{c}{$n=200$} 
& &\multicolumn{4}{c}{$n=500$} \\
\cline{3-6} \cline{8-11} 
&&Model A &Model B&Model C&Model D
&&
Model A &Model B&Model C&Model D
\\
\hline
$\gamma_0$& Bias& 
-0.0977& -0.0499& -0.0977&  -0.3087
& & -0.0129&   0.0187&  -0.0415&-0.0277\\
&ASE&   0.5709&  0.3678&0.6963& 0.5964
& &  0.3556& 0.2335& 0.3475&  0.2756\\
&SD&  0.4392& 0.3169&0.6216& 0.4400
& &  0.2913&  0.2109& 0.3998&0.2813\\
&CP& 0.9640& 0.9670&  1.0000&  1.0000
& & 0.9740&  0.9750&  0.9000& 0.9670\\
$\gamma_1$& Bias& 
 -0.0528&  0.0278& 0.1228&  0.1174
& &   -0.0401&  -0.0192& -0.0693&-0.0156\\
&ASE&  0.7316& 0.5344&  0.8152&  0.6905
& &   0.4400& 0.3106&  0.4541& 0.3437\\
&SD&0.6985& 0.5281& 0.7973&  0.5073
& & 0.4140& 0.3024&  0.5052& 0.3409\\
&CP& 0.9880& 0.9730&  1.0000&  1.0000
& & 0.9810&  0.9680&  0.9430& 0.9930\\
$\gamma_2$& Bias& 
 -0.1844&  -0.1313& 0.3491&  0.1725
& & -0.0774&  -0.0703&   0.2346&0.0774\\
&ASE&    0.3554&   0.2986&  0.5027&  0.3586
& & 0.2123&  0.1845&  0.3138& 0.1700\\
&SD& 0.3608&  0.2837& 0.3401&  0.3178
& &0.2212&   0.1923& 0.4024&0.1989\\
&CP&0.9780& 0.9600& 1.0000&  0.9330
& & 0.9640&  0.9560&  0.9290& 0.9200\\
$\eta_0$& Bias& 
 -0.0696&  -0.0351&  -0.4513& -0.4524
& &  -0.0282&-0.0118&  -0.2212&-0.0369\\
&ASE& 0.7769&  0.6036&  0.9577& 0.6623
& &0.4688&  0.3609&  0.6787& 0.5600\\
&SD&0.4911& 0.4562&  1.0266& 0.4610
& & 0.3439&  0.3039&  0.6118& 0.4603\\
&CP&0.9760& 0.9670& 0.8460& 0.7330
& &0.9730&   0.9740&   0.9000& 0.9270\\
$\eta_1$& Bias& 
-0.3212& -0.2229&  0.1771& 0.0279
& & -0.1015&  -0.0790&  0.0888&-0.0494\\
&ASE& 0.6758&0.5064& 0.4863&  0.3443
& & 0.3710&   0.2864&  0.2795&0.2500\\
&SD&0.7956& 0.5167&  0.7767& 0.6419
& &0.3330&  0.2957&   0.3331& 0.2554\\
&CP&0.9720& 0.9530& 0.9230&  0.8000
& & 0.9750& 0.9560&  0.9290&0.9400\\
$\eta_2$& Bias& 
0.2505&  0.1728&  0.2085&  0.2387
& & 0.0749&  0.0501& 0.1796&   0.1763\\
&ASE&0.4779&  0.3666&   0.4713&    0.5487
& & 0.2406&   0.1968& 0.3687& 0.4061\\
&SD& 0.7633&  0.4198&  0.4941&  0.6899
& &  0.2498&  0.2115&  0.6013& 0.5336\\
&CP& 0.9520& 0.9530&  1.0000&   1.0000
& &0.9620&   0.9600&   0.8430& 0.9400\\
$\eta_3$& Bias& 
-0.0883&-0.1231&   0.1522&  0.0849
& & -0.0452&   -0.0352&  -0.0254&0.0024\\
&ASE&0.7309& 0.6037&  0.8035&  0.5923
& &  0.4150&  0.3464&  0.4941&0.4174\\
&SD&0.7270&  0.5315&  0.9964& 0.5120
& &  0.4040&  0.3566&0.6547&  0.3854\\
&CP&0.9640& 0.9800&  0.9230&  1.0000
& &  0.9600&   0.9560&  0.9000&0.9530\\
$\eta_4$& Bias& 
 0.2745&  0.1592&  0.1301&  0.3628
& &  0.1083&  0.0600&  0.1119& 0.0122\\
&ASE&0.8343&0.6607&  1.0210&  1.0211
& & 0.4570& 0.3647&  0.7216&0.6472\\
&SD& 0.9021& 0.7038&   1.0031&  1.4017
& &0.4571&    0.3607&  0.8808& 0.6544\\
&CP&0.9700& 0.9400&  0.9200& 0.9330
& &0.9540&   0.9560&  0.9000 & 0.9800\\
\hline
	\end{tabular}
\end{table}

\clearpage
\begin{table}[h!]
\caption{Simulation results for the case is both $X$ and $Z$ are  vectors of covariates}  
\label{tab:case4} \centering \tabcolsep=3.5pt 
\begin{tabular}{cl cccc c cccc }
\hline
& & \multicolumn{4}{c}{$n=1000$} 
& &\multicolumn{4}{c}{$n=2000$} \\
\cline{3-6} \cline{8-11} 
&&Model A &Model B&Model C&Model D
&&
Model A &Model B&Model C&Model D
\\
\hline
$\gamma_0$& Bias& 
-0.0009& 0.0102&  -0.0459&  -0.0474
& & -0.0016& 0.1111&   -0.0208&-0.0087\\
&ASE&    0.2480&  -0.1608&  0.2479&  0.2019
& &  0.1729&  0.1111&  0.1747& 0.1440\\
&SD&  0.2167& 0.1505&   0.2707&   0.2091
& &0.1665&   0.1096&  0.1938& 0.1471\\
&CP& 0.9710& 0.9660&  0.9330&  0.9420
& & 0.9580& 0.9520&   0.9080& 0.9480\\
$\gamma_1$& Bias& 
0.0058&  0.0033&  -0.0072&  0.0281
& &  0.0019&  0.0076&  0.0042& 0.0009\\
&ASE&  0.3066&   0.2134&  0.2911
&  0.2367
& & 0.2121&  0.1470& 0.1929& 0.1620\\
&SD&0.3085& 0.2106&  0.3182&  0.2478
& & 0.2189& 0.1559&   0.2105&0.1676\\
&CP&  0.9590& 0.9690& 0.9570&  0.9530
& & 0.9520&   0.9490&  0.9480& 0.9560\\
$\gamma_2$& Bias& 
-0.0324&   -0.0250& 0.0941& 0.0491
& & -0.0145&   -0.0129&   0.0371& 0.0166\\
&ASE&   0.1467&  0.1259&   0.1811& 0.1221
& & 0.1002& 0.0860&   0.1209&0.0855\\
&SD&0.1414&  0.1313& 0.2324& 0.1377
& & 0.1014&  0.0870&  0.1465&0.0908\\
&CP& 0.9670& 0.9510&  0.9260&  0.8980
& & 0.9530&  0.9560&   0.8920&0.9180\\
$\eta_0$& Bias& 
-0.0056&  0.0036&  -0.0413&   -0.0577
& & 0.0175&  0.0179&  0.0071& 0.0176\\
&ASE&  0.3287&  0.2562&   0.5453&  0.3809
& & 0.2321&  0.1813& 0.4061&0.2961\\
&SD& 0.2749&  0.2287&   0.5761& 0.3480
& & 0.2178&  0.1782&  0.4519& 0.2795\\
&CP&  0.9710&  0.9740&   0.8890&   0.9040
& & 0.9660&  0.9640& 0.9000& 0.9440\\
$\eta_1$& Bias& 
-0.0444&  -0.0379&    0.0479&  -0.0304
& & -0.0216&  -0.0194&   0.0362& -0.0210\\
&ASE&  0.2522&   0.1993&   0.2125&   0.1718
& & 0.1756&   0.1392&  0.1566&0.1271\\
&SD&0.2355&  0.1902& 0.2443&  0.1849
& & 0.1711&   0.1369& 0.1766& 0.1317\\
&CP& 0.9710&  0.9620&  0.9250&  0.9190
& &  0.9570& 0.9540&   0.9110& 0.9280\\
$\eta_2$& Bias& 
0.0306&  0.0191&  0.0759&  0.0821
& &  0.0043&  0.0039&  0.0595& 0.0421\\
&ASE& 0.1600&  0.1335& 0.2811&0.2503
& &  0.1103&    0.0933&  0.2041&0.1796\\
&SD&0.1638&  0.1354&   0.3301&  0.3195
& & 0.1124& 0.0927& 0.2269& 0.2028\\
&CP& 0.9440& 0.9510&  0.8850&  0.9040
& & 0.9490&   0.9550&  0.8990& 0.9150\\
$\eta_3$& Bias& 
-0.0230&  -0.0148& -0.0655&  -0.0104
& &-0.0168&  -0.0151&  -0.0467&-0.0270\\
&ASE& 0.2831&  0.2403& 0.3463&  0.2756
& & 0.1980&  0.1684&   0.2444& 0.1976\\
&SD& 0.2675&  0.2363 & 0.3839&  0.2792
& & 0.1902&  0.1646&  0.2534&0.1961\\
&CP& 0.9700& 0.9600&  0.9590&  0.9650
& &  0.9530&  0.9510&   0.9470&  0.9580\\
$\eta_4$& Bias& 
0.0221& 0.0058&  0.0132& 0.0765
& & 0.0042&  0.0015&  0.0223& 0.0080\\
&ASE& 0.3114&   0.2526& 0.5331& 0.4169
& & 0.2176& 0.1770& 0.3784&0.2930\\
&SD&0.3110&  0.2438&   0.5664&   0.4610
& & 0.2176& 0.1768&  0.4010&0.2980\\
&CP& 0.9570&0.9600& 0.9580& 0.9540
& & 0.9570&  0.9560&  0.9540& 0.9640\\
		\hline
	\end{tabular}
\end{table}

\section{A practical example}
To properly do regression analysis for real data sets, it requires performing five steps like the following algorithm.
\subsection{Algorithm}
\begin{itemize}
\item [] Step 1: Check attributes and features of the actual data set to identify the exact data set (count or ZI count data, binary or ZI binary data, and so on).
\item [] Step 2: From step 1, select appropriate regression models to investigate the actual data set.
\item [] Step 3: Based on the selection model criteria, to select the most appropriate regression model for the actual data set under consideration.
\item [] Step 4: Use some popular calculation software such as R or Matlab to find the estimated parameters of the regression model.
\item [] Step 5: Draw findings from these results.
\end{itemize}

\subsection{The fishing data set}
\begin{figure}[h]
\begin{center}
\includegraphics[width=140mm]{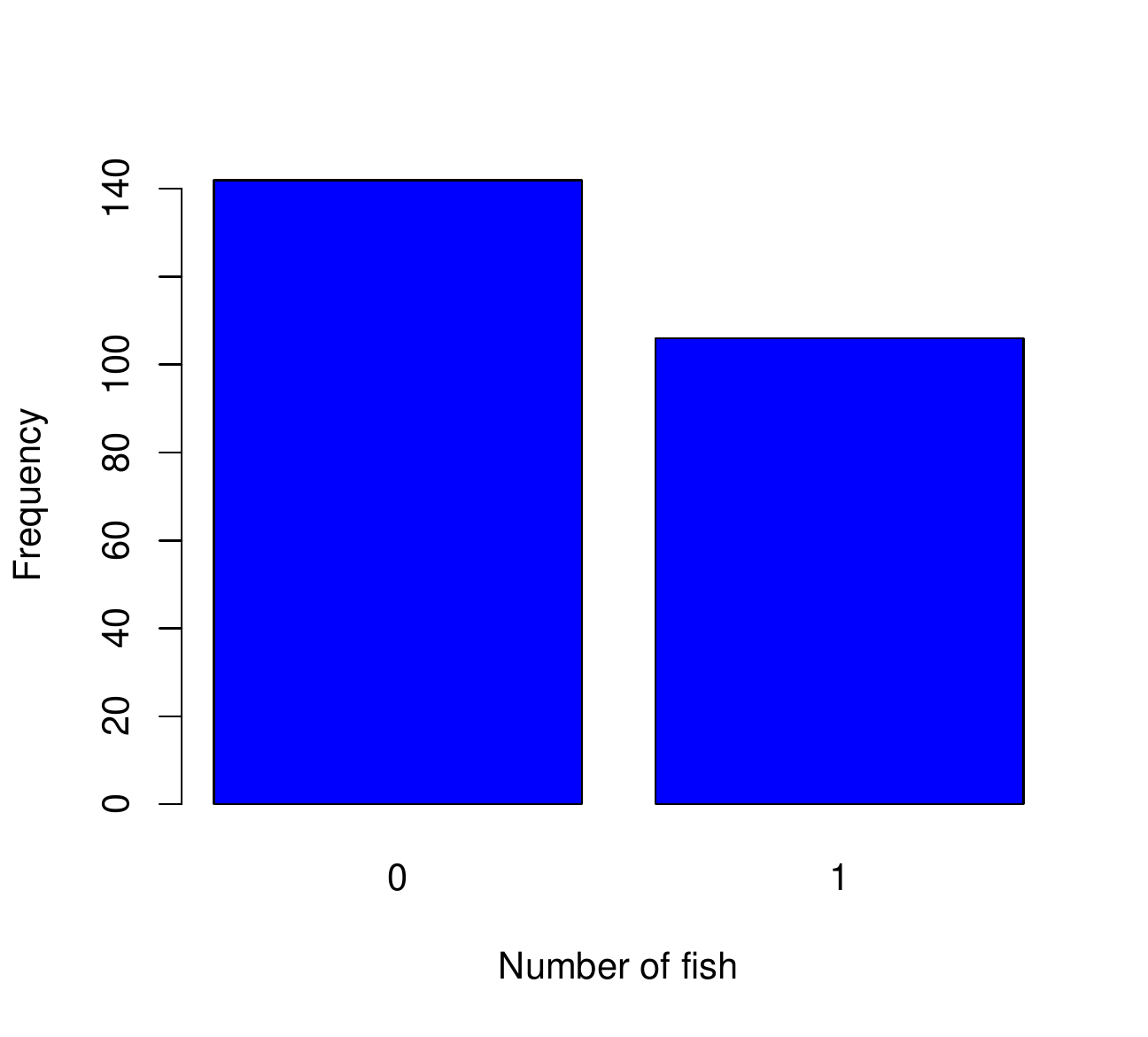}
\end{center}
\caption{Bar plot of the number of fish. \label{plot-fish}} 
\end{figure}

The fishing data set (FDS) was used to test the performance of the models considered in this work. The FDS was first introduced and analyzed in the studies of Li and Lu (2021) and Pho and McAleer (2021). A total of 248 subjects were surveyed in the FDS. Each group or subject was asked several questions such as: the number of fish they caught ($\textbf{fish\_caught}$), do they use livebait or not? ($\textbf{livebait}$), did they bring campers to the park? ($\textbf{camper}$), how many people in the group go fishing? ($\textbf{persons}$), how many children are in the group ($\textbf{children}$) and how long they stay ($\textbf{hours}$). Note that the ratio of zeros in the FDS is 142/248=0.5726 (see Figure \ref{plot-fish}) suggests that the response is zero-inflated. In addition, this is a binary data set, so it can be fully named ZI binary data (Step 1). We have introduced and considered four regression models in this study: logit-ZIBer, probit-ZIBer, cloglog-ZIBer and GEV-ZIBer. Indeed, all of these regression models can be used for regression analysis for the FDS. To know which model will be more suitable for regression analysis for the FDS, we rely on the Vuong test (see Pho et al., 2019). 

In order to choose the most suitable regression model for regression analysis for the FDS, we also first try to test the models under consideration in this study with the two most popular models in the binary models family: probit and logit.
For the sake of simplicity we choose the probit-ZIBer model as the standard model and consider all other models (Step 2). We first use the Vuong test to check the compatibility of probit-ZIBer and the probit and logit model. The results in these cases are 7.73 and 7.43 respectively. It can be seen that these numbers are all greater than 1.96, so the probit-ZIBer model will be more suitable for the FDS in regression analysis than the two familiar binary models, probit and logit.

We also next use the probit-ZIBer model to compare its compatibility for the FDS in regression analysis with the logit-ZIber, clogclog-ZIBer and GEV-ZIBer models. The results obtained in these situations are 6.66, 31.56 and 210.53, respectively. Based on the evidences from the Vuong test, because these figures are all greater than 1.96, it can be seen that the probit-ZIBer model will be the most suitable for regression analysis for the FDS (Step 3). We therefore use the probit-ZIBer model to perform the regression analysis for the FDS with the outcome variable considered as the number of fish they caught ($\textbf{fish\_caught}$). In addition, we aim to use two predictors $\textbf{persons}$ (how many people in the group go fishing?) and $\textbf{livebait}$ (do they use livebait or not?) as variables X and Z respectively.

The probit-ZIBer used in this empirical analysis has the following form.
\begin{align*}
P(Y=1|X,Z)=
\frac{e^{\gamma_0 + \gamma_1Z}}{1+e^{\gamma_0 + \gamma_1Z}}\frac{e^{\eta_0 + \eta_1X + \eta_2Z}}
{1+e^{\eta_0 + \eta_1X + \eta_2Z}}.
\end{align*}
Similar to the simulation studies, we also use R statistical software to obtain the results in the experimental analysis. The results of this experimental analysis are detailed in Table 5 (Step 4).

\begin{table}[h!]
\caption{Results of the experimental analysis.}
\label{tab:surveydata1}
\begin{center}
\begin{tabular}{cc  ccc}	  
\hline
Variable&Parameter&$\widehat{\bm\beta}$& ASE & $p$-value \\ 
\hline  \\ [-1.5mm]
\multicolumn{5}{l}{\textit{\textbf{Model 1}} }\\ [0.8mm]
Intercept &$\gamma_0$&  -0.2598&  0.0809&$<0.0013$\\
$Z$&$\gamma_1$& ~0.0826& 0.0187&$<0.0001$\\
[0.8mm]
\multicolumn{5}{l}{\textit{\textbf{Model 2}} }\\ [0.8mm]
Intercept&$\eta_0$& -1.2612& 0.1637&$<0.0001$\\
$X$ &$\eta_1$& ~2.4117&  0.3693&$<0.0001$\\
$Z$&$\eta_2$& ~0.6660& 0.0606
&$<0.0001$\\[0.8mm]
\hline 
\end{tabular}
\end{center}
\end{table}
In general, the values obtained for the p-value are very small, all are less than 5\% and therefore all the estimates for the regression parameters in this empirical analysis are statistically significant. We next give some comments for the variables of interest in this study (Step 5).

First comment on the variable X, the variable that refers to $\textbf{persons}$ and the estimated result obtained for this variable is 2.4117. It is easy to see that this is a positive number, which also implies that if we increase this variable by some amount, the result obtained for the outcome variable will also increase accordingly. In other words, it also means that if the number of people in this variable is increased, the number of fish caught while fishing also increases. This implication is very good, has a deep meaning and is true to reality. It helps us to increase economic efficiency or revenue in fishing.

In the following, we also make a few comments for the other explanatory variable, that is, the variable Z, which represents the $\textbf{livebait}$, with values of 0.0826 and 0.6660, respectively. This will also have the same implication as the variable X, which means that if we increase this variable by some amount, the output of the resulting variable will also increase. Put plainly, if we increase by some amount the variable Z, which represents the livebait, then the number of fish caught during fishing will also increase. If we pay more attention and comment in more detail, the estimate of variable X will be larger than the estimate of variable Z. And this will mean that if we aim to increase by one given a certain amount for the explanatory variables, the effect of increasing some amount for variable X will be higher than increasing by some amount for variable Z. This implication is very interesting and makes a lot of sense in practice.

\section{ Concluding and perspectives}
We examined the stability and robustness of the parameter estimation in the ZIBer model, when the SP model is modeled by several different binary models: logit, probit, cloglog and GEV.  The MLE approach is used to check its performance when different SP models are considered. According to the numerical illustrations through simulation studies and the analysis of a real data set, it has been seen that the MLE method has provided accurate and reliable inferences. Besides, it can also be seen that for the empirical analysis, the probit-ZIBer model is probably more suitable for the fishing data set than the other models considered in this work. Moreover, the results obtained in the actual analysis are also very consistent, compatible and very meaningful in practice. This will help us to understand the importance of increasing production while fishing.

In the remainder of this study, we would like to point out that some studies regarding the ZIBer model can be considered in future researches. For example, an extension of the ZIBer model which allows for the response variable to be measured with error may be developed. Note that 
the ZIBer model which allows for the explanatory variable to be measured with error is developed in the work of Lee, Pho and Li (2021). Additionally, one may introduce the ZIBer model with random effects. Moreover, the topic of goodness of fit for the ZIBer model with missing covariates has also not been conducted.
Another issue of interest deals with the inference the Bayesian approach for parameter estimation of the ZIBer model and so on.

\clearpage
\section*{References}
\def\hang{\hangindent=\parindent\noindent}
\hang
Ali, E. (2022), A simulation-based study of ZIP regression with various zero-inflated submodels. \textit{Communications in Statistics-Simulation and Computation}, 1-16.

\hang
Bodromurti, W., Notodiputro, K. A., and Kurnia, A. (2018), Zero Inflated Binomial Model for Infant Mortality Data in Indonesia, \textit{International Journal of Applied Engineering Research}, 13(6), 3139-3143.

\hang
Cox, D.R. (1958), The regression analysis of binary sequences, \textit{Journal of the Royal Statistical Society: Series B (Methodological)}, 20(2), 215-232.

\hang
Diallo, A.O., Diop, A., and Dupuy, J.F. (2017), Asymptotic properties of the maximum-likelihood estimator in zero-inflated binomial regression, \textit{Communications in Statistics-Theory and Methods}, 46(20), 9930-9948.

\hang
Diallo, A.O., Diop, A., and Dupuy, J.F. (2019), Estimation in zero-inflated binomial regression with missing covariates, \textit{Statistics}, 53(4), 839-865.

\hang
Diop, A., Diop, A., and Dupuy, J.F. (2011), Maximum likelihood estimation in the logistic regression model with a cure fraction, \textit{Electronic journal of statistics}, 5, 460-483.

\hang
Diop, A., Diop, A., and Dupuy, J.F. (2016), Simulation-based inference in a zero-inflated Bernoulli regression model, \textit{Communications in Statistics-Simulation and Computation}, 45(10), 3597-3614.

\hang
Hall, D.B. (2000), Zero-inflated Poisson and binomial regression with random effects: a case study, \textit{Biometrics}, 56(4), 1030-1039.

\hang
He, H., Wang, W., Hu, J., Gallop, R., Crits-Christoph, P., and Xia, Y. (2015), Distribution-free inference of zero-inflated binomial data for longitudinal studies, \textit{Journal of applied statistics}, 42(10), 2203-2219.

\hang
Lee, S. M., Pho, K. H., and Li, C. S. (2021), Validation likelihood estimation method for a zero-inflated Bernoulli regression model with missing covariates. \textit{Journal of Statistical Planning and Inference}, 214, 105-127.

\hang
Li, C. S., and Lu, M. (2021), Semiparametric zero-inflated Bernoulli regression with applications, \textit{Journal of Applied Statistics}, 1-25.

\hang
Pho, K. H. (2022), Goodness of fit test for a zero-inflated Bernoulli regression model. \textit{Communications in Statistics-Simulation and Computation}, 1-16.

\hang
Pho, K. H., Ly, S., Ly, S., and Lukusa, T. M. (2019), Comparison among Akaike information criterion, Bayesian information criterion and Vuong's test in model selection: A case study of violated speed regulation in Taiwan, \textit{Journal of Advanced Engineering and Computation}, 3(1), 293-303.

\hang
Pho, K.H., and McAleer, M. (2021), Specification and estimation of a logistic function, with applications in the sciences and social sciences, \textit{Advances in Decision Sciences}, 25(2), 1-31.

\hang
Rakitzis, A.C., Maravelakis, P.E., and Castagliola, P. (2016), CUSUM control charts for the monitoring of zero-inflated binomial processes, \textit{Quality and Reliability Engineering International}, 32(2), 465-483.

\hang
Truong, B. C., Pho, K. H., Dinh, C. C., and McAleer, M. (2021), Zero-inflated poisson regression models: Applications in the sciences and social sciences. \textit{Annals of Financial Economics}, 16(02), 2150006.

\hang
Wang, X. and Dey, D. K. (2010), Generalized extreme value regression for binary response data:An application to B2B electronic payments system adoption.
 \textit{ Ann. Appl. Stat.} 4 2000-2023.

\hang
Wanitjirattikal, P., and Shi, C. (2020), A Bayesian zero-inflated binomial regression and its application in dose-finding study, \textit{Journal of Biopharmaceutical Statistics}, 30(2), 322-333.

\end{document}